\documentclass{JINST}

\title{A $^{83}$Kr$^m$ Source for Use in Low-background Liquid Xenon Time Projection Chambers}

\author{L.~W.~Kastens, S.~Bedikian, S.~B.~Cahn, A.~Manzur, and D.~N.~McKinsey   \llap{$^a$}
\\
\llap{$^a$}Department of Physics, Yale University, P.O. Box 208120, New Haven, CT 06520.}

\abstract{We report the testing of a charcoal-based $^{83}$Kr$^m$ source for use in calibrating a low-background two-phase liquid xenon detector. $^{83}$Kr$^m$ atoms produced through the decay of $\rm ^{83}Rb$ are introduced into a xenon detector by flowing xenon gas past the $\rm ^{83}Rb$ source.  9.4~keV and 32.1~keV transitions from decaying $^{83}$Kr$^m$ nuclei are detected through liquid xenon scintillation and ionization.  The characteristics of the $^{83}$Kr$^m$ source are analyzed and shown to be appropriate for a low background liquid xenon detector.  Introduction of $^{83}$Kr$^m$ allows for quick, periodic calibration of low background noble liquid detectors at low energy.}

\begin{document}

\section{Introduction}
Cosmological evidence points towards the existence of dark matter, which makes up 23\% of the mass-energy density of the universe \protect\cite{WMAP}.  A
strong candidate for dark matter is the WIMP (Weakly Interacting
Massive Particle), which is the subject of many direct dark matter
searches\protect\cite{GoodmanWitten, Jungman, Bertone, Gaitskell}, in which
elastic nuclear scatters present a potential WIMP signal. Liquefied noble gases have recently become 
important materials for direct searches for WIMPs, with several
experiments achieving strong WIMP-nucleon cross-section limits over
the past few years\protect\cite{Angle,Alner,Lebedenko,LArTPCs}. 

\begin{figure}[htb]
\centering
\includegraphics[width=0.6\textwidth, angle=0]{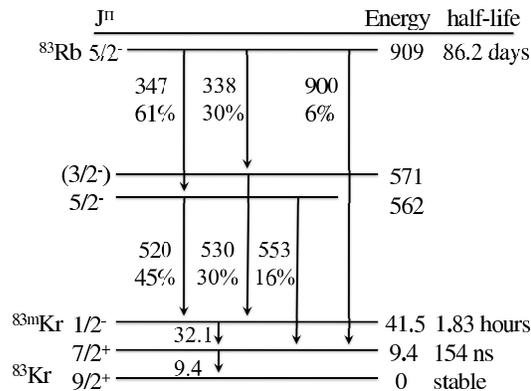}
\caption[Label]{Energy level diagram (in keV) for the $^{83}$Rb decay.  $^{83}$Rb 
decays
75$\%$ of the time
to the long-lived isomeric $^{83}$Kr$^m$ level that is 41.5~keV above the ground state, which
subsequently decays in two steps, a 32.1~keV transition, typically a conversion electron and associated x-rays, followed by a similar 9.4~keV transition~\protect\cite{Kr83mInZeolite}.} \label{Rb83}
\end{figure}

Calibration of noble liquid detectors at low energies (tens of keV) has become a challenge as
detector sizes have increased.  The self-shielding capability of
large detectors, which is greatly advantageous for reaching low radioactive background levels, prevents gamma rays from external calibration sources from penetrating to the fiducial volume.  Two techniques have been explored for the doping of a low-energy radioactive source into the active region of noble liquid detectors to allow for the efficient production of low-energy
events in the fiducial volume.  XENON10 used two activated xenon isotopes produced at Yale
University\protect\cite{ActivatedXe} and shipped to Gran Sasso National Laboratory to calibrate the fiducial volume.  The isotopes $^{129}$Xe$^m$ and $^{131}$Xe$^m$ emit 236 keV and
164 keV gammas with half-lives of 8.9 and 11.8 days, respectively. 
While these isotopes can be used to fill the detector with a temporary
calibration source, the energies are higher than those expected from
WIMP signals, and their long half-lives prevent their use for frequent
calibration. A second such source is derived from ${^{83}}$Rb, which
decays to $^{83}$Kr$^m$ with a half-life of 86.2 days. The
$^{83}$Kr$^m$ subsequently decays via 32.1 keV and
9.4 keV transitions with a half-life of 1.83
hours~\protect\cite{Kr83mInZeolite} (Figure~\protect\ref{Rb83}). In this paper, we describe tests of this source for use in a low-background dark matter experiment, verifying the chemical purity, radiopurity, and suitability for periodic calibration. Measurements of $^{83}$Kr$^m$ were previously made using a single phase liquid xenon detector by Kastens \textit{et~al}~\protect\cite{ourpaper}.  More recently, dual phase liquid xenon measurements were described in Manalaysay \textit{et~al}~\protect\cite{Zurich}.  Calibration of liquid argon and liquid neon detectors with $^{83}$Kr$^m$ is described in Lippincott \textit{et~al}~\protect\cite{hugh}.

\section{Experimental Apparatus}

\begin{figure}[htbp]
\centering
\includegraphics[width=0.6\textwidth ]{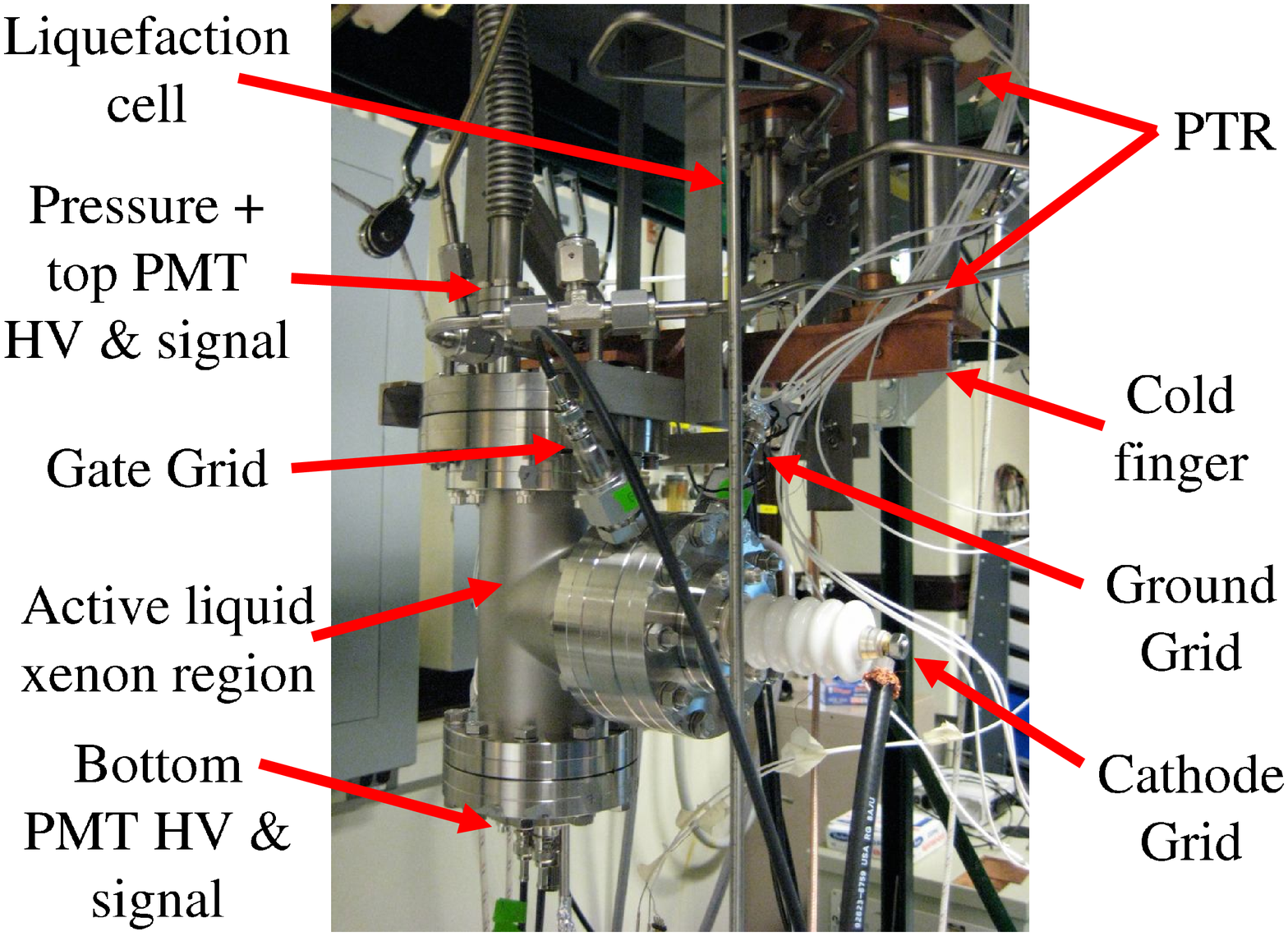}
\caption{Photograph of apparatus for $^{83}$Kr$^m$ measurement in liquid xenon.}
\label{pic}
\includegraphics[width=0.6\textwidth ]{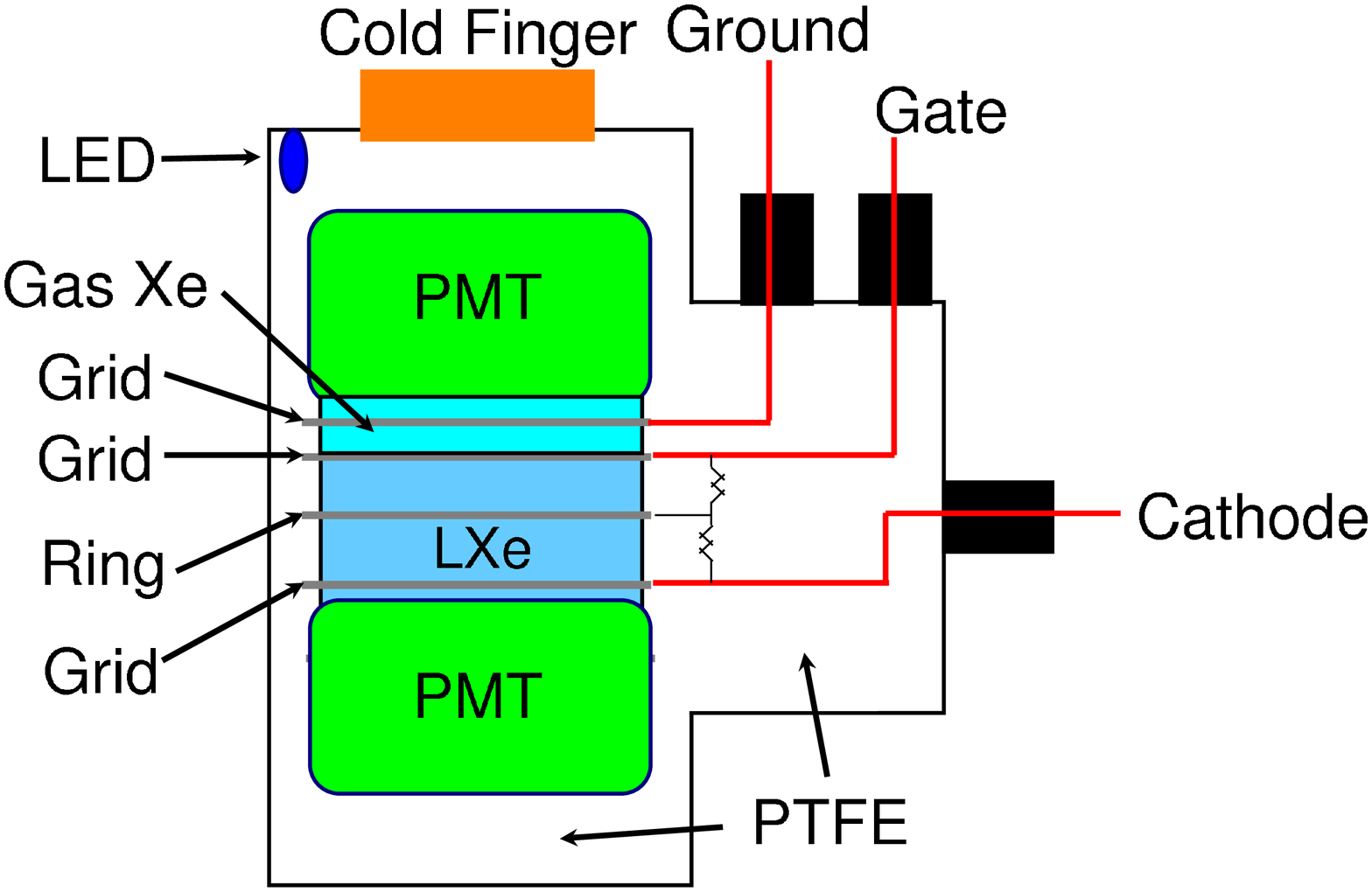}
\caption{Schematic of apparatus for $^{83}$Kr$^m$ measurement in liquid xenon.  A field shaping ring helps create a uniform electric field between the cathode and the gate grid to drift charge through the liquid.  An electric field between the gate grid and the ground grid drift charge through the gaseous xenon and produces proportional scintillation light from electrons extracted from the liquid xenon.  }
\label{setup}
\end{figure}

The two-phase liquid xenon detector used to test $^{83}$Kr$^m$ calibration is
located at Yale
University (Fig \ref{pic}). The two-phase xenon detector has an active mass of 150~g, with an active volume 5~cm in diameter and 2~cm in height, viewed by two photomultiplier tubes (PMTs).  It has been previously described in \protect\cite{angel}, while the cryogenic and gas handling systems are described in \protect\cite{ActivatedXe}.  Stainless steel wire grids apply a drift field of 1.0~kV/cm in the liquid, and a proportional scintillation field of 8.0~kV/cm in the gas.  

The gains of the two PMTs are measured from the single photoelectron
(phe) spectra by using light emitted from an LED inside the liquid xenon detector.  The PMT bias was subsequently lowered to prevent PMT saturation from the proportional scintillation pulse.  The liquid xenon detector is then calibrated with 122~keV and 136~keV gamma rays from a $^{57}$Co source outside the cryostat.  The gain was scaled by the shift in the $^{83}$Kr$^m$ scintillation peak after the PMT bias was changed.  The scintillation signal yield for the $^{57}$Co gamma rays in liquid xenon is measured to be 4.3~phe/keV at a drift field of 1.0~kV/cm. 

\begin{figure}[htbp]
\centering
\includegraphics[width=0.65\textwidth]{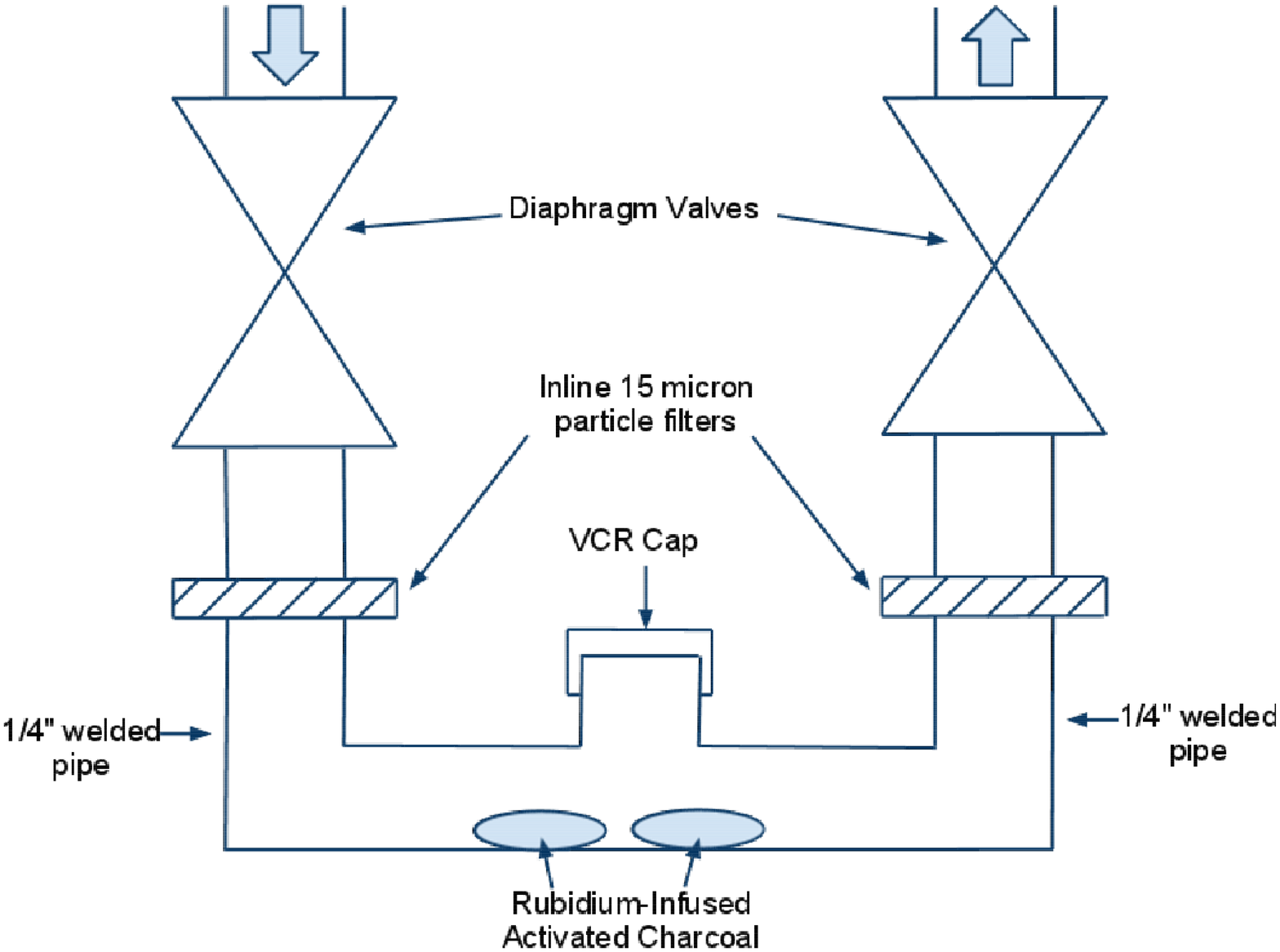}
\caption{A schematic of the $^{83}$Kr$^m$ generator.  The valve, in-line particle filters, and tee are all 1/4" components.  The tee closed by the VCR cap is the only access point to the interior of the generator.  All other joints are welded.  Rubidium-infused activated charcoal pellets are introduced via this tee, which is then sealed.  }
\includegraphics[width=0.65\textwidth, angle=0]{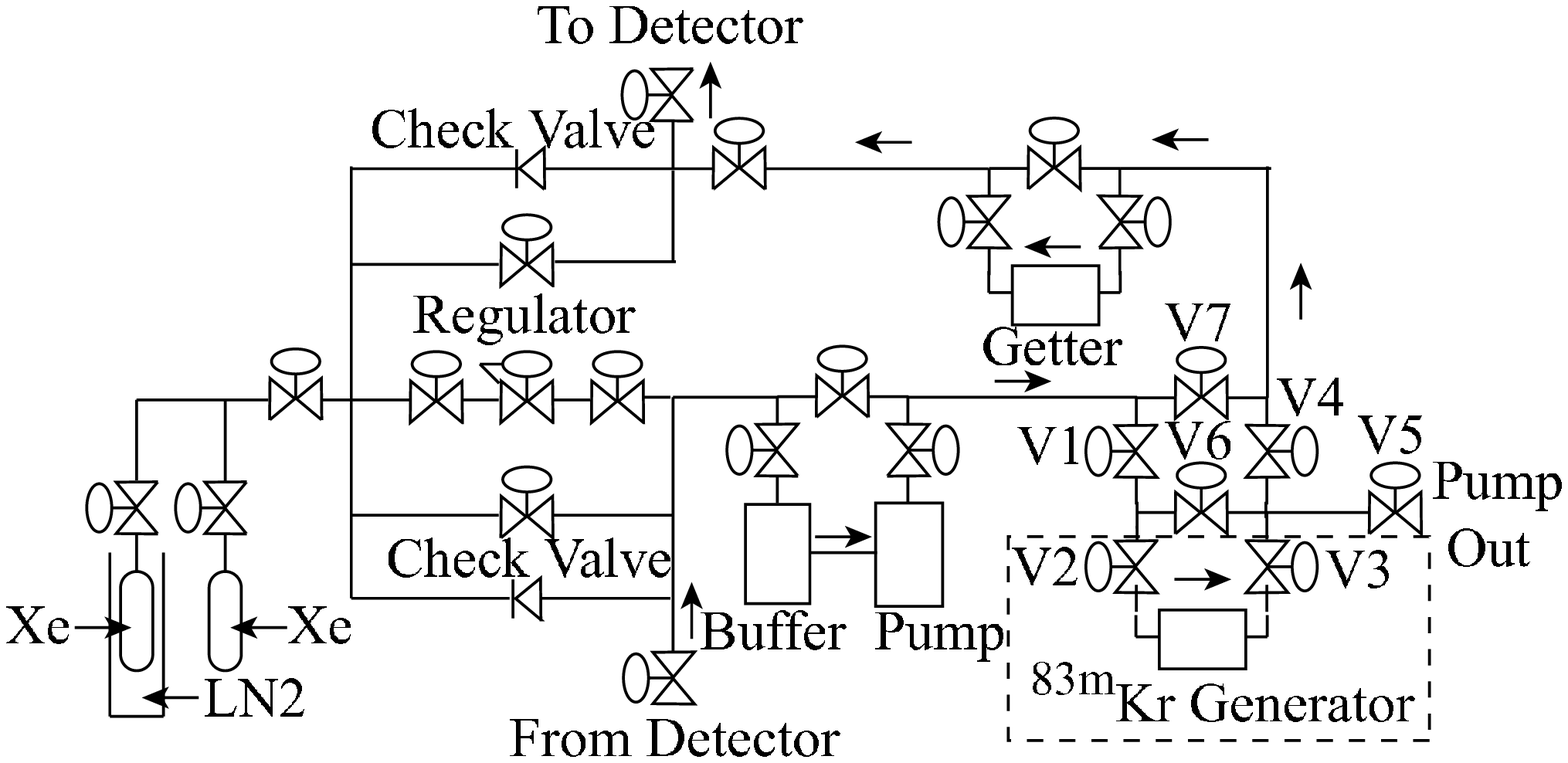}
\caption{A diagram of the attachment of the $^{83}$Kr$^m$ generator to the xenon gas handling system.  Xenon is typically circulated through V7 to maintain detector purity.  $^{83}$Kr$^m$ is introduced by opening V1, V2, V3 and V4, while closing V7.  }
\label{fig:generator}
\end{figure}

A diagram of the $^{83}$Kr$^m$ generator is in figure \ref{fig:generator}.  400~nCi of $^{83}$Rb from the RbHCl solution described in \protect\cite{ourpaper} was adsorbed into two pieces of activated charcoal and was introduced to the tee through the VCR arm into the space between the two particulate filters.  The arm of the tee was then sealed and the generator purified via baking and pumping.  

Possible emission of rubidium from the generator was tested using a residual gas analyzer (RGA).  Natural rubidium (0.1~mL aqueous RbCl solution) was applied to a sample of charcoal.  After absorption of the rubidium, the pellets were placed in a vacuum chamber while a RGA monitored the level of rubidium escaping into the vacuum.  There was no natural rubidium detected above the baseline of the RGA (10$^{-14}$ A in ion current or fewer than 1 in 10$^{11}$ atoms detected were rubidium). This placed an upper limit on the unintended egress of 8000~Rb atoms per second.  Only 20~$\mu$L of radioactive RbCl solution was introduced to the $^{83}$Kr$^m$ generator, placing an upper limit on the unintended egress from the $^{83}$Kr$^m$ generator at 3~Rb atoms per month.  A limit of 3~Cl atoms per month is obtained in the same fashion.  

Potential radioactive contamination is only from chemicals added to standard high-purity plumbing: the rubidium and chlorine used to load the generator, and the radon emitted from the charcoal.  The remainder of the generator consists of standard high-purity plumbing, whose potential contribution to radiopurity is negligible compared to the rest of the xenon plumbing system.  The only long-lived, gaseous, radioactive isotope produced from rubidium and chlorine contamination will be $^{81}$Kr, which emits a 276~keV in less than 1\% of decays with a 230,000~year halflife, and will not be a significant background.  The appearance of other radioactive peaks was looked for after the experiments described here, but none were found.  

Based on uranium gamma ray measurements from the SOLO germanium detector \protect\cite{SOLO}, the zeolite mediator described in \protect\cite{ourpaper} produced radon at a rate of 45,000~atoms per day.  To meet background goals, such an apparatus used on the LUX (Large Underground Xenon) detector, a dark matter direct detection experiment scheduled to be deployed to SUSEL (Sanford Underground Science and Engineering Laboratory), would require a radon emission rate of less than 1,400~radon atoms per day for a one-day exposure to the detector, assuming that beta decays due to radon daughters occur isotropically inside the liquid xenon volume of the detector.  This limit is set by constrains on the allowable rate of $^{214}$Bi beta decay, which can produce low-energy events in the fiducial volume.  

The experiment described here uses 20~grams of an activated coconut carbon mediator from Calgon (OVC4x8).  Activated carbon binds strongly to radon, and has been used to purify gaseous nitrogen, liquid nitrogen, and air for low background experiments \protect\cite{borex, poncthesis}.  Radon emanation from this type of charcoal has been measured to be approximately 9.4~mBq/kg \cite{poncthesis}, or less than 0.5~mBq for a 20 gram sample after the generator is initially pumped out.  This corresponds to a Radon generation rate of less than 41~radon atoms per day, less than the limit of 1400~radon atoms per day for LUX.  

\section{Data Analysis and Results}

Events were recorded via a 250~MHz, 12-bit CAEN V1720 waveform digitizer for later analysis. Data acquisition was triggered by coincident PMT pulses, each greater than 30~phe. The digitized PMT waveforms were analyzed to quantify the response of the detector to $^{83}$Kr$^m$ decays. 

\begin{figure}[htb]
\centering
\includegraphics[width=0.6\textwidth ]{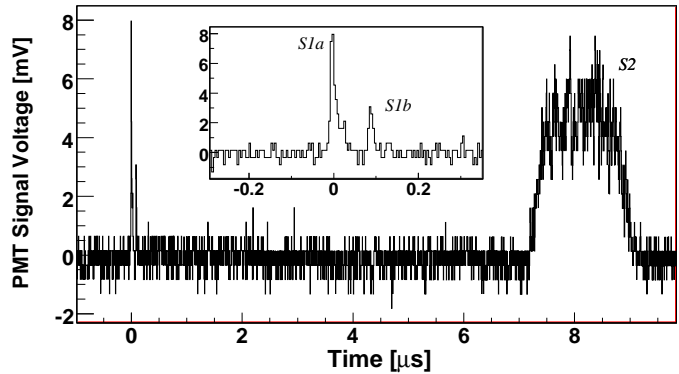}
\caption[Label]{A typical $^{83}$Kr$^m$ event as seen by the bottom PMT.  The main plot shows both the prompt scintillation (\textit{S1a} and \textit{S1b}) and proportional scintillation (\textit{S2}) pulses, while the inset zooms in on the prompt scintillation pulses.  The drift time between \textit{S1a} and \textit{S2} indicates the depth at which the event occurred in the TPC.  After the electrons are drifted through the liquid and extracted into the gas, each electron generates 23.2$\pm$0.7 proportional scintillation photoelectrons, creating the much larger proportional scintillation signal. } 
\label{fig:event}
\end{figure}

A typical $^{83}$Kr$^m$ waveform contains three pulses (Figure \protect\ref{fig:event}). The first pulse, \textit{S1a}, corresponds to the prompt scintillation due to the 32.1~keV transition. The second pulse, \textit{S1b}, corresponds to the prompt scintillation due to the 9.4~keV transition, emitted following the first transition with a 154~ns half-life.  The third pulse (\textit{S2}) is a combined 41.5~keV proportional scintillation signal generated by electrons associated with both transitions and drifted through the liquid.  The 32.1~keV and 9.4~keV \textit{S2} pulses cannot be distinguished as the \textit{S2} width is much larger than the time between the \textit{S1} pulses.  Each pulse was integrated, then divided by the gain of the PMT to determine the number of photoelectrons detected. The number of photoelectrons in each PMT was added and the sum divided by the signal yield determined from the $^{57}$Co calibration. 

\begin{figure}[htb]
\centering
\includegraphics[width=0.6 \textwidth ]{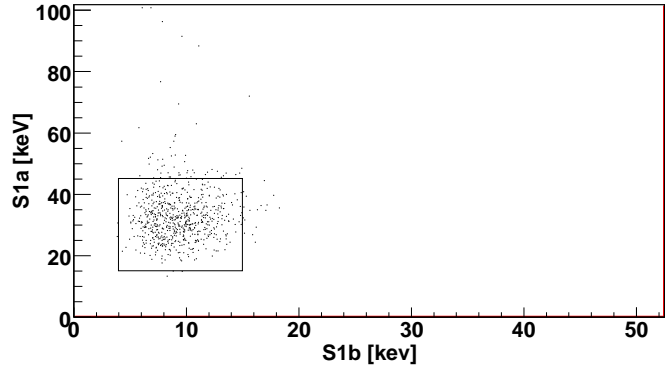}
\caption[Label]{A scatter plot of \textit{S1a} and \textit{S1b}.  The box indicates the acceptance region, in which \textit{S1a} is required to be between 15~keV and 45~keV, \textit{S1b} is required to be between 4~keV and 15~keV, and \textit{S2} is required to be between 2000~phe and 6000~phe.  Furthermore, \textit{S1b} must occur within 400~ns of \textit{S1a}.  Before $^{83}$Kr$^m$ was introduced, the event rate was less than $6.0\times10^{-4}$~Bq in this region.  After introduction, the event rate in this region rose to $6.5\times10^{-1}$~Bq in this region.  } 
\label{fig:S1aS2}
\end{figure}

A demonstration of the introduction of $^{83}$Kr$^m$ into the liquid xenon volume using a charcoal mediator is effected through a scatter plot of \textit{S1a} and \textit{S1b} after flowing xenon gas through the $^{83}$Kr$^m$ generator at 2~liters per minute.  Standard cuts for identifying $^{83}$Kr$^m$ events are as follows: \textit{S1a} must be between 15~keV and 45~keV, \textit{S1b} must be between 4~keV and 15~keV, \textit{S2} must be between 2000~phe and 6000~phe, and \textit{S1b} must occur within 400~ns of \textit{S1a}.  Additional standard cuts require the \textit{S1a} asymmetry (defined as $(\textit{S1}_{top}-\textit{S1}_{bot})/(\textit{S1}_{top}+\textit{S1}_{bot})$) to be between -0.9 and -0.65, the \textit{S1b} asymmetry to be between -0.9 and -0.65, the \textit{S2} pulse width to be above 400~ns, and the electron drift time to be between 1.0~$\mu$s and 10.0~$\mu$s.  These cuts ensure that these events occured with proper pulse characteristics for \textit{S1} and \textit{S2} pulses.  Figure \protect\ref{fig:S1aS2} shows a scatter plot of \textit{S1a} vs. \textit{S1b} using these standard cuts.  Before the introduction of $^{83}$Kr$^m$, the event rate was less than $6.0\times10^{-4}$~Bq.  After introduction, the event rate rose to $6.5\times10^{-1}$~Bq.  
A demonstration of the introduction of $^{83}$Kr$^m$ into the liquid xenon volume using a charcoal mediator is effected through a scatter plot of \textit{S1a} and \textit{S1b} after flowing xenon gas through the $^{83}$Kr$^m$ generator at 2~liters per minute.  Standard cuts for identifying $^{83}$Kr$^m$ events are as follows: \textit{S1a} must be between 15~keV and 45~keV, \textit{S1b} must be between 4~keV and 15~keV, \textit{S2} must be between 2000~phe and 6000~phe, and \textit{S1b} must occur within 400~ns of \textit{S1a}.  Additional standard cuts require the \textit{S1a} asymmetry (defined as $(\textit{S1}_{top}-\textit{S1}_{bot})/(\textit{S1}_{top}+\textit{S1}_{bot})$) to be between -0.9 and -0.65, the \textit{S1b} asymmetry to be between -0.9 and -0.65, the \textit{S2} pulse width to be above 400~ns, and the electron drift time to be between 1.0~$\mu$s and 10.0~$\mu$s.  These cuts ensure that these events occured with proper pulse characteristics for \textit{S1} and \textit{S2} pulses.  Figure \protect\ref{fig:S1aS2} shows a scatter plot of \textit{S1a} vs. \textit{S1b} using these standard cuts.  Before the introduction of $^{83}$Kr$^m$, the event rate was less than $6.0\times10^{-4}$~Bq.  After introduction, the event rate rose to $6.5\times10^{-1}$~Bq.  

\begin{figure}[htb]
\centering
\includegraphics[width=0.6\textwidth ]{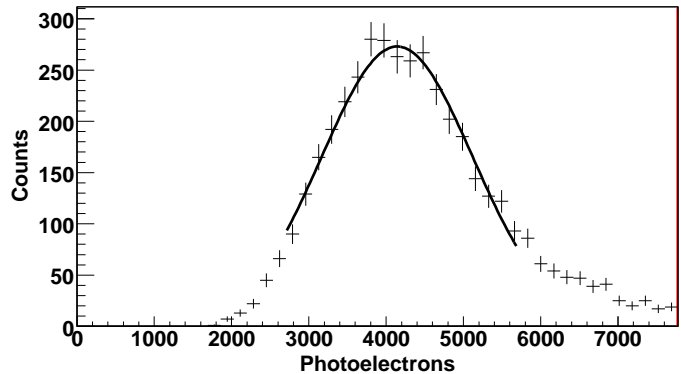}
\caption[Label]{A plot of the $^{83}$Kr$^m$ 41.5~keV proportional scintillation pulse, in triple coincidence with the two primary scintillation pulses.  Quality cuts are also made on the \textit{S2} pulse asymmetry and timing width.  The peak has a resolution (\textit{$\sigma$/E}) of 23\%. } 
\label{fig:peaks}
\end{figure}

The \textit{S2} pulse, proportional to the charge signal, is plotted in Figure \protect\ref{fig:peaks} using the standard cuts.  The mean, corresponding to a 41.5~keV ionization pulse, is at 4,143$\pm$22~phe with a resolution ($\sigma/E$) of 23\%.  

\begin{figure}[htb]
\centering
\includegraphics[width=0.6\textwidth ]{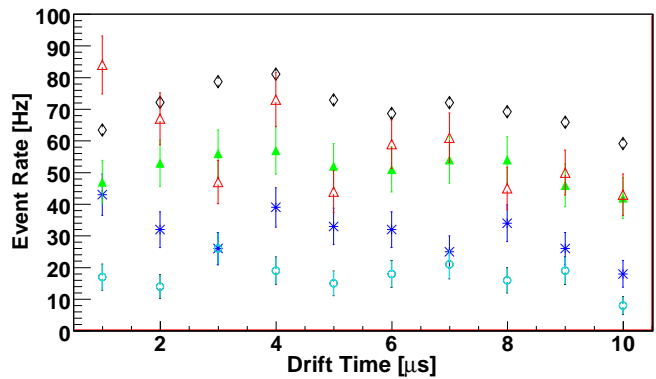}
\caption[Label]{The distribution of events after cuts in drift time as the krypton is introduced into the detector.  
Data is from 60~s ($\circ$), 360~s ($\ast$), 600~s ($\blacktriangle$) and 900~s ($\triangle$) after introduction of $^{83}$Kr$^m$.  The ($\diamondsuit$) points represent the steady state distribution after the generator has been open to the detector for over 4~hours.  The steady state distribution is constant to within 12\%.  The other distributions are consistent with the steady state distribution scaled by a constant.  } 
\label{fig:z_dist}
\end{figure}

The distribution of events in drift time may be determined by the time between \textit{S1a} and \textit{S2}.  The distribution of events in drift time as $^{83}$Kr$^m$ was introduced to the system is shown in Figure \protect\ref{fig:z_dist}.  The standard cuts were used, with the exception of the PMT asymmetry cuts which affect the drift time distribution of events.  Each curve is the drift time distribution at various times as the $^{83}$Kr$^m$ was introduced into the detector.  Each distribution is consistent with the steady state rate, which is uniform to within 12\%, scaled by a constant.  This suggests that once the $^{83}$Kr$^m$ is introduced into the active volume, $^{83}$Kr$^m$ rapidly becomes uniformly distributed over the height (2~cm) of this detector. 

\begin{figure*}[htb]
\centering
\includegraphics[width=0.9\textwidth ]{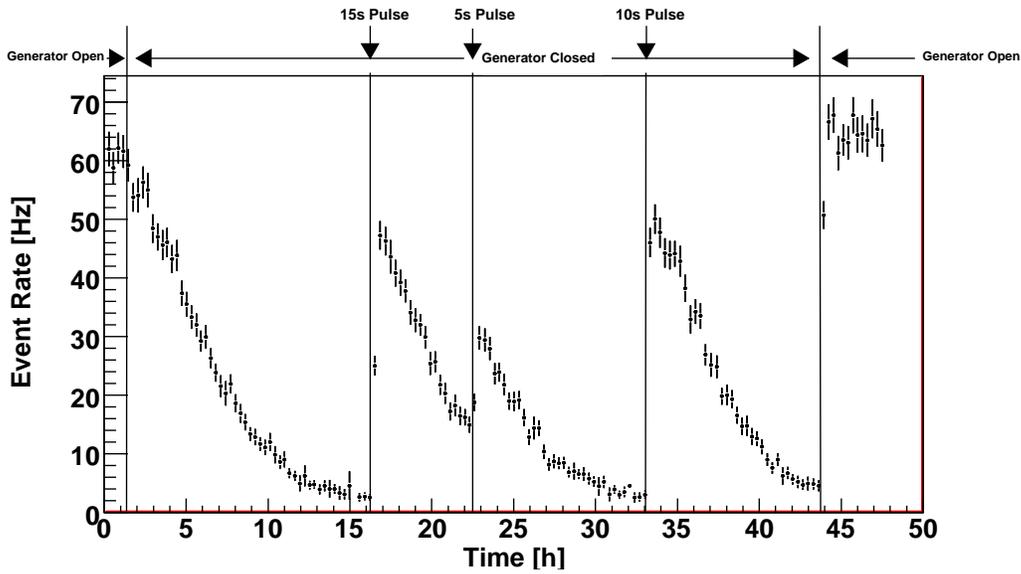}
\caption[Label]{The rate of $^{83}$Kr$^m$ as a function of time.  The generator began open to the detector for a two days before observation began.  The generator was closed after 1.4~h.  A 15~s pulse of $^{83}$Kr$^m$ was introduced after 16.25~h, a 5~s pulse after 22.5~h, and a 10~s pulse after 33.1~h.  After 43.7~h, the $^{83}$Kr$^m$ generator was opened to the detector.  } 
\label{fig:rate}
\end{figure*}

Figure \protect\ref{fig:rate} shows the $^{83}$Kr$^m$ rate as a function of time using the standard cuts.  Pulses for 5~s, 10~s, and 15~s demonstrate the ability to easily control the $^{83}$Kr$^m$ rate through actuation of valves surrounding the generator.  The latency between introducing a pulse and observing the $^{83}$Kr$^m$ decay is approximately 300~s, while the maximum $^{83}$Kr$^m$ rate is achieved in under 900~s.  

\section{Discussion}
The purity of the $^{83}$Kr$^m$ generator is of utmost importance for low-background experiments.  In particular, no long-lived radioisotopes should be introduced with the $^{83}$Kr$^m$, and the generator should be easily purified to avoid a loss in charge collection when $^{83}$Kr$^m$ is introduced.  After construction of the generator, xenon gas was continuously circulated through the generator and a getter to remove any impurities inside the generator.  After purification of the $^{83}$Kr$^m$ generator, no significant change in electron lifetime was observed over 48 hours as multiple pulses of $^{83}$Kr$^m$ were introduced to the detector, corresponding to a change in impurity level (oxygen-equivalent) of less than 0.4~ppm.  A change of impurity level of 0.4~ppm in a 2~kg detector corresponds to a reduction of the drift length of less than 1\% in a 350~kg detector with a 50~cm drift length.

After flowing xenon through the $^{83}$Kr$^m$ generator for over two weeks, the detector was then isolated from the generator.  After a week, the rate of $^{83}$Kr$^m$ in the detector had reached equilibrium with any potential $^{83}$Rb present in the detector volume.  The rate of $^{83}$Kr$^m$ was measured to be less than 0.72~mBq, or 36~nBq for every 60~s generator exposure.  

The 36~nBq $^{83}$Rb rate per minute of generator exposure is three orders of magnitude below the background requirement for LUX for a total of 1000~minute exposure to the $^{83}$Kr$^m$ source.  Only a trace amount of radioactive $^{36}$Cl is present in the environment.  Based on the emission of chlorine as measured by the RGA, less than one radioactive chlorine atom will be released every $10^{9}$~years.  

The ability to introduce $^{83}$Kr$^m$ in under 5 minutes allows for frequent introduction of $^{83}$Kr$^m$ to calibrate the energy scale in the fiducial region, map detector scintillation and ionization response, and monitor detector stability.  External calibration sources are limited as most of the events occur outside the fiducial volume due to self-shielding and saturate the PMTs, reducing the rate at which calibration data can be collected as detector size grows.  This source could replace periodic external gamma ray calibrations, both reducing calibration time and providing a large number of events in the fiducial volume and energy region of interest.  

The $^{83}$Kr$^m$ source will not necessarily provide a uniform decay rate throughout a large liquid xenon detector.  While we observe the distribution to be roughly uniform on short (2~cm) scales, the mixing of $^{83}$Kr$^m$ in xenon could be much less over larger distances, causing different rates of $^{83}$Kr$^m$ decays at different points in the detector.  However, such a source does not need to be perfectly uniform as long as sufficient statistics can be acquired at each portion of the detector.  A non-uniform source still allows mapping of the scintillation and ionization response of the detector, though it would not permit this calibration method to be used to test x-y position reconstruction.  

The $^{83}$Kr$^m$ mixing will allow measurements of fluid flow and turbulence during purification.  Xenon purification can occur faster than 1/$e$ per volume change if mixing is not present.  A $^{83}$Kr$^m$ source could be used to find mixing and non-mixing modes of fluid flow.  Ideally, a mixing mode of purification could be used in conjunction with the $^{83}$Kr$^m$ source to test the robustness of x-y position reconstructions, while a non-mixing mode could be used when speed of purification is a primary concern.  

\section{Summary}
We have demonstrated the ability to introduce and detect $^{83}$Kr$^m$ in a two phase xenon detector using a low-background $^{83}$Kr$^m$ generator.  Two scintillation pulses from 9.4~keV and 32.1~keV transitions were detected, followed by a combined proportional scintillation signal.  The ability to control the source over time was demonstrated, and the spatial distribution of events over time was measured.  

The generator allows efficient introduction of $^{83}$Kr$^m$ to the detector, allowing calibration on demand.  Generators relying on diffusion \protect\cite{Zurich} to introduce the $^{83}$Kr$^m$ to the detector take hours instead of minutes to introduce the $^{83}$Kr$^m$ to the active region.  This is disadvantageous when attempting to accumulate livetime on a low background experiment such as a dark matter search.  Flowing xenon gas through the generator allows quick introduction of $^{83}$Kr$^m$, facilitating periodic calibrations with a minimum of calibration time.

\section{Acknowledgments}
The authors would like to recognize the assistance of Kevin Charbonneau in the acquisition of the $^{83}$Rb and preparation of the $^{83}$Rb-infused charcoal, and David Malling for facilitating the uranium and thorium measurements of the zeolite at SOLO.  This work was supported by NSF Grant PHY-0800526.

\end{document}